\documentstyle[aps,epsf]{revtex}

\begin{document}
%\draft
\twocolumn
\title{ Collapse and revival of ultracold atoms in a microwave cavity
and of photons in parametric down-conversion}
\author{Karl-Peter Marzlin}
\address{School of Mathematics, Physics, Computing and Electronics, 
	Macquarie University, Sydney, NSW 2109, Australia.}
\author{J\"urgen Audretsch}
\address{Fakult\"at f\"ur Physik der Universit\"at Konstanz,
	Postfach 5560 M 673, D-78434 Konstanz, Germany}
\maketitle
%%%%%%%%%%%%%%%%%%%%%%%%%%%%%%%%%%%%%%%%%%%%%%%%%%%%%%%%%%%%%%%%
\begin{abstract}
We present a new theoretical method to study a trapped gas of
bosonic two-level atoms interacting with a single mode of a microwave
cavity. This interaction is described by a trilinear Hamiltonian which
is formally completely equivalent to the one describing parametric
down-conversion in quantum optics. A system of differential
equations describing the evolution, including the long-time
behaviour, of not only the mean value but also the variance of
the number of excited atoms is derived and solved analytically.
For different initial states the mean number of
excited atoms exhibits periodically reappearing dips, with an
accompanying peak in the variance, or fractional collapses and
revivals. Closed expressions for the period and the revival time
are obtained.
\end{abstract}
%%%%%%%%%%%%%%%%%%%%%%%%%%%%%%%%%%%%%%%%%%%%%%%%%%%%%%%%%%%%%%%%
$ $ \\
%\narrowtext
%%%%%%%%%%%%%%%%%%%%%%%%%%%%%%%%%%%%%%%%%%%%%%%%%%%%%%%%%%%%%%%
\section{The physical systems}
Since the experimental realization of weakly interacting atomic
Bose-Einstein condensates confined in magnetic traps
\cite{exp} much work was done to study the interaction of such a
system with light. Examples include the scattering of light by
a condensate \cite{you96}, its refraction index \cite{morice95},
nonlinear atom optics \cite{lenz94,zhang94}, the optical detection
of the relative phase between two condensates \cite{java96},
and the spontaneous emission in the presence of two condensates
\cite{walls97}.

In this paper we are concerned with the question of how a condensate
of two-level atoms interacts with a single resonant
quantized cavity mode of the electromagnetic field. The system can
be realized by placing the trap containing the condensate in a
microwave cavity. We consider the limit of vanishing temperature
so that essentially all atoms are in the condensate.
The two states $|g \rangle$ and $|e \rangle$ are both members of
the internal ground-state manifold of the condensed atoms but
have a different hyperfine magnetic quantum number $m_F$.
For both states
$m_F$ should be chosen in such a way that the atoms remain
trapped after the transition from $|g \rangle$ to $|e\rangle$.
The internal energies corresponding to these states are denoted
by $E_g$ and $E_e$, respectively. Although for the existing
condensates the trapping potential
is different for states having a different magnetic quantum number
we assume for simplicity that the trapping potential is
the same for $|g \rangle$ and $|e \rangle$. This situation may
be realized by using non-magnetic traps or by exposing the atoms
to a highly detuned laser beam. If the shape of the beam is
appropriately chosen the induced dipole-potential can compensate
for the difference between the trap potentials.

Under these conditions the Hamiltonian in its second quantized
form is given by
\begin{eqnarray} 
  H_0 &=& \int d^3x \sum_{i=e,g} \Big \{ \Psi_i^\dagger ( \vec{x}) 
  \Big [H_{c.m.}  + E_i \Big ] \Psi_i (\vec{ x}) \Big \}
  \nonumber \\ & &
  +\hbar \omega_c a^\dagger a + H_{int} + H_{n.l.}
\label{start} \end{eqnarray}
where $\Psi_i$, $i=e,g$ is the field operator for atoms in the
state $|g\rangle$ and $|e\rangle$, respectively. 
$a$ is the
annihilation operator for photons in the microwave cavity. 
The center-of-mass Hamiltonian
\begin{equation} 
  H_{c.m.} := \frac{ -\hbar^2
  \Delta }{2M} + {M\over 2} \omega_z^2 z^2  + {M\over2}
  \omega_\perp^2 (x^2+y^2)
\end{equation} 
is the same for both internal states. $M$ denotes the atomic mass,
and $\omega_\perp$ and $\omega_z$ are the trap frequencies in
the $x-y$ plane and in the z-direction, respectively. 
The nonlinear Hamiltonian
\begin{equation} 
  H_{n.l.} = 
  {1\over2} \int d^3 x \sum_{i,j=e,g}
  g_{ij}\Psi_i^\dagger (\vec{ x})\Psi_j^\dagger (\vec{x}) 
  \Psi_j(\vec{x}) \Psi_i (\vec{ x})  \; .
\end{equation}
describes the interaction between the atoms. The quantities
$g_{ij}$ are connected to the scattering lengths for the
scattering between atoms in the internal states $|i\rangle$
and $|j\rangle$. We will make the Hartree approximation
and consider the case that all atoms are in the same
state $\varphi_0(\vec{x})$ regarding the center-of-mass motion.
This state fulfills the stationary nonlinear Schr\"odinger equation
(see, e.g., Ref. \cite{nlse}) with energy eigenvalue $\mu$
and is the same for both internal states.

We model the interaction of the atoms with the microwave photons
by the usual magnetic dipole coupling
\begin{equation} 
  H_{int} = -[a
  \vec{{\cal B}}  + a^\dagger \vec{{\cal B}}^* ]
  \cdot \int d^3 x \{ \vec{m}_{ge}
  \Psi_e^\dagger (\vec{ x})\Psi_g (\vec{x}) + H.c. \}
\end{equation}
where $\vec{m}_{ge}$ is the magnetic dipole moment of the atoms and
$\vec{{\cal B}}$ the magnetic field of the cavity mode at the
position of the trap. As we consider the case that the trap size
is much smaller than the wavelength of the microwave photons
the interaction does not affect the center-of-mass motion of the
condensed atoms. This justifies the assumption that the atoms will
always remain in the spatial mode $\varphi_0 (\vec{x})$.

Under these assumptions we can neglect all spatial modes beside
$\varphi_0(\vec{x})$ and can replace the atomic field operators
$\Psi_i(\vec{x})$ by $\varphi_0(\vec{x})
b_i (\vec{x})$, $i=e,g$, where the operators $b_i :=
\int d^3 x \varphi_0^*(\vec{x}) \Psi_i(\vec{x})$ are the
annihilation operators for atoms in the internal state
$|i \rangle$. Substituting this in the Hamiltonian (\ref{start})
the latter can be simplified to
\begin{eqnarray} H &=& (\mu + E_e)b_e^\dagger b_e +
  (\mu +E_g)b_g^\dagger b_g + \hbar \omega_c a^\dagger a
  \nonumber \\ & & - [a
  \vec{{\cal B}}  + a^\dagger \vec{{\cal B}}^* ]
  \cdot
  \{ b_g^\dagger b_e \vec{m}_{ge}  + b_g b_e^\dagger
  \vec{m}_{ge}^*\} \; .
\end{eqnarray}
In the interaction picture and after the rotating
wave approximation the Hamiltonian
reduces in resonance ($\hbar \omega_c = E_e-E_g$) to
\begin{equation} \tilde{H} = - \hbar \Omega \{ b_g^\dagger b_e
a^\dagger + b_g b_e^\dagger a \} \; .
\label{htilde}\end{equation}
The Rabi frequency $\Omega$ is given by $| \vec{m}_{ge}\cdot
\vec{{\cal B}}^*|/\hbar$.

The Hamiltonian (\ref{htilde}) is formally equivalent to that used
to describe parametric down conversion in nonlinear
optics (see, e.g., Ref.~\cite{mandel95}). In this system $b_e$
destroys a photon of the pumping beam and $a^\dagger$ and $b_g^\dagger$
create a signal and an idler photon, respectively. The interaction between
the different modes is caused by a nonlinear medium and describes
the process that a pumping photon is converted into two other photons
under conservation of energy. Theoretically the pumping
beam is often considered as a classical field. To be short
we will focus in the following on the interpretation of the
Hamiltonian (\ref{htilde})
as an interaction between two-level atoms in a cavity. Though
the physical interpretation is different our results can
immediately be applied to parametric down conversion if one
considers $N_e = b_e^\dagger b_e$
as the number operator for pumping photons.

A trilinear Hamiltonian has been examined numerically by
Walls and Barakat \cite{walls70}. Kumar and Mehta \cite{kumar80}
derived analytical expressions for the time evolution
under the condition that the number of one of the three particle
species involved in Eq.~(\ref{htilde}) remains strongly populated. 
That the degenerate parametric
oscillator may show collapse and revival has been predicted numerically by
Jyotsna and Agarwal \cite{iyyanki97}. These results are similar to the
numerical results of Drobn\'y and Jex \cite{drobny92}.
An exact solution for a fixed number of atoms 
was given in a non-closed form by Tavis and
Cummings \cite{tavis69}.
%%%%%%%%%%%%%%%%%%%%%%%%%%%%%%%%%%%%%%%%%%%%%%%%%%%%%%%%%%%%%%%%%%%
\section{Coupled differential equations for the mean occupation
number and its variance}
Our aim is to present a new analytical method for solving the
underlying operator equations. In particular this will then enable us
to describe for the system given above
the long-time behaviour of the mean occupation $\bar{N}_e(\tau)$
of the excited atomic mode for the case that initially
all atoms are excited and there are zero photons in the cavity
and to confront it with the time dependence of its variance.
To do so we want to give closed-form approximate analytical solutions for
$\bar{N}_e(\tau)$ for different initial states of the condensate like atomic
number state, coherent state, or mixture. This will allow us to demonstrate
that $\bar{N}_e(\tau)$ shows a periodic behaviour or the existence of
collapse and revival, respectively.

The Hamiltonian (\ref{htilde}) leads to two
conserved quantities: the total number of atoms $S_A :=
b_e^\dagger b_e + b_g^\dagger b_g$ and, because of the
rotating-wave approximation, the number of excitations $S_E :=
b_e^\dagger b_e + a^\dagger a$. Because any two of the
number operators $N_e, N_g, N_a$ can be expressed by the third and
the two  conserved quantities, it suffices to solve the Heisenberg
equation for only one of the three number operators. We will consider
the number operator $N_e$ of excited atoms.

Introducing the new
time variable $\tau :=  \Omega t$ and considering the Heisenberg
equation for $N_e$ as well as for
\begin{equation} \dot{N}_e = -i b_g^\dagger b_e
a^\dagger +i b_g b_e^\dagger a \; ,
\label{ndot}\end{equation}
where the derivative is taken with
respect to $\tau$, one easily finds (see, e.g., Ref.~\cite{kumar80})
\begin{equation} \ddot{N}_e = 6 N_e^2 - 2 A N_e + 2 B\; .
\label{2ndorder}\end{equation}
For notational convenience we have introduced the operators
$A :=  2S_E+ 2S_A + 1$ and $B:= S_A S_E$. Note that
while $S_A$ and $S_E$ commute with every operator occurring in the problem,
$\dot{N}_e$ does not commute with $N_e$. This fact seems to prevent
the integration of the second order differential
equation to an equation of first order as it is usually done
with differential equations for ordinary functions.
A common way to circumvent this difficulty is the {\em vanishing variance
approximation}, where a differential
equation for $\bar{N}_e$ is
derived under the assumption that the variance $\Delta_e$ defined by
$\Delta_e^2 := \langle (N_e - \bar{N_e})^2
\rangle$ is negligible.
Here and in the following the mean value of an operator $O$ will be
represented by a bar, $\bar{O} := \langle \psi | O |\psi \rangle$.
The complete state of the system is denoted by $|\psi \rangle$.
For the system under consideration this approximation was studied
by Kumar and Mehta \cite{kumar80} starting from Eq.~(\ref{2ndorder}).
They showed that it is valid for short times only
and can be applied only to a very limited class of physical problems.

Our objective is to improve this scheme:
We want to go a step further and derive differential equations for
the mean value $ \bar{N}_e $ as well as for the variance $\Delta_e$
and study the respective long time behaviour
for different states of the condensate. Because of
$\partial_{\tau}^2 \langle N_e^2 \rangle = \langle \ddot{N}_e
N_e + N_e \ddot{N}_e + 2 \dot{N}_e^2 \rangle$, the
discussion of $\Delta_e(\tau)$ needs the
knowledge of $\dot{N}_e^2$ and therefore the exact integration of the
operator equation (\ref{2ndorder}). This can indeed be done by observing
that the commutator between $N_e$ and $\dot{N}_e$
is given by $ [N_e,\dot{N}_e] = \tilde{H}/(i\hbar \Omega)$ so that
$N_e^2 \dot{N}_e + \dot{N}_e N_e^2 = (2/3) (d N_e^3 /d \tau) + (1/3)
\dot{N}_e $ follows, what can be used to integrate Eq.~(\ref{2ndorder}) to
\begin{equation} (\dot{N}_e)^2 = 4 N_e^3 - 2A N_e^2
	+ 2 N_e (2B+1)+C\; . \label{1stdgl} \end{equation}
Here $C$ is a constant of motion which after some algebra can be
written as $C = -\tilde{H}^2/(\hbar^2 \Omega^2) + 2S_A S_E$.
It is therefore related to the conservation of energy and particle
numbers. Note that no approximation has been made up to this point.

Although Eq.~(\ref{1stdgl}) cannot be exactly integrated because $C$ does
not commute with $N_e$, we are now able to treat $\bar{N}_e$ together with
$\Delta_e$ if we make the two following assumptions
of a {\em vanishing asymmetry approximation}:
(i) the conserved particle numbers are always uncorrelated to the
number of excited atoms so that $\langle S_i N_e^l \rangle =
\langle S_i \rangle \langle N_e^l \rangle $ holds for $i=A,E$ and any
integer $l$. This equality is exactly fulfilled for all times if the
initial state is an eigenstate of $S_i$, e.g., a number state.
(ii) we can always neglect $\langle (N_e -\bar{N}_e)^3 \rangle$
so  that $\langle N_e^3 \rangle$ can be approximated by
$\bar{N}_e^3 + 3 \bar{N}_e \Delta_e^2$. Accordingly, in contrast to the
vanishing variance approximation, we only assume the
vanishing of the higher moment $\langle (N_e -\bar{N}_e)^3 \rangle$
in the hierarchy of moments $\langle (N_e -\bar{N}_e)^m \rangle , m=2,3,
\ldots $ of the probability distribution. The case $m=3$
describes its degree of asymmetry.

Based on this we derive from
Eqs. (\ref{2ndorder}) and (\ref{1stdgl}) the system of ordinary
differential equations
\begin{eqnarray}
\partial_{\tau}^2 \bar{N}_e &=& 6(\Delta_e^2 + \bar{N}_e^2) -2
	\bar{A} \bar{N}_e +2\bar{B}\; ,
	\label{sys1} \\
  \partial_{\tau}^2 \Delta_e^2 &=& 20 \bar{N}_e^3 +60 \Delta_e^2 \bar{N}_e -8
  \bar{A} (\Delta_e^2 + \bar{N}_e^2)
  \nonumber \\ & &
  +4(1+ 3
   \bar{B}) \bar{N}_e +2 \bar{C} - \partial_{\tau}^2
   \bar{N}_e^2 \; ,\label{sys2}
\end{eqnarray}Eq.~(\ref{sys1}) may be considered as an algebraic equation for $\Delta_e^2$
which can be inserted into Eq.~(\ref{sys2}) to decouple the system. This
results finally in
\begin{eqnarray}
\partial_{\tau}^4 \bar{N}_e &=& - 10 \bar{A}
	\partial_{\tau}^2 \bar{N}_e + 60 \bar{N}_e
	\partial_{\tau}^2 \bar{N}_e -240 \bar{N}_e^3
	+120 \bar{A} \bar{N}_e^2
	 \nonumber \\ & &
	+\bar{N}_e (24- 48 \bar{B}-16 \bar{A}^2)
	+12 \bar{C} + 16\bar{A}\bar{B}
	\label{nosys}
\end{eqnarray}
$\Delta_e$ can then be obtained from Eq.~(\ref{sys1}).
%%%%%%%%%%%%%%%%%%%%%%%%%%%%%%%%%%%%%%%%%%%%%%%%%%%%%%%%%%%%%%%%%%%%
\section{Analytical solution for a number state}
We are now able to treat the envisaged physical situation: all atoms are
initially for $\tau=0$ excited and no photons are present. It is
desirable to obtain the development in time of $\bar{N}_e(\tau)$ for the
case of the atomic state being a number state with $n$ atoms, the
corresponding complete state of the system being denoted by $| \psi_n\rangle$.
We will demonstrate below that $\bar{N}_e(\tau)$ for other physical situations
can be reduced to this.
The operator equations (\ref{2ndorder}) and (\ref{1stdgl}) show that the
initial conditions for $\bar{N}_e(\tau)$ which are to be fulfilled for a
number state are: $\bar{N}_e(0) =n$, $ \partial_\tau \bar{N}_e(0) =
\partial_\tau^3 \bar{N}_e(0) =0$, and $\partial_\tau^2 \bar{N}_e(0) = -2n$.

We have in fact been able to find an exact solution of
Eq.~(\ref{nosys}) fulfilling the correct number state initial conditions for
$\bar{N}_e(0), \partial_\tau \bar{N}_e(0), \partial_\tau^3 \bar{N}_e(0)$, and,
to leading order in $n$, for $\partial_\tau^2 \bar{N}_e(0)$.
It is given by
\begin{equation}
\bar{N}_e^{\mbox{\scriptsize num}}(\tau)= \bar{N}_e^{\mbox{\scriptsize
	num}}(0) - {1\over 2} m \omega^2\mbox{cn}^2 (\omega \tau
	- K(m)| m)\; ,
\label{lsg}\end{equation}
where the parameters $m$ and $\omega$ are solutions of the algebraic
equations
\begin{eqnarray}
0 &=& \omega^4 m(m-1)\{5  [6 \bar{N}_e^{\mbox{\scriptsize
	num}}(0)- \bar{A}] -2 \omega^2(2m-1)\}
      \nonumber \\ & &
      - 120 (\bar{N}_e^{\mbox{\scriptsize num}}(0))^3
      + 60 \bar{A} (\bar{N}_e^{\mbox{\scriptsize num}}(0))^2
      + 8 \bar{A} \bar{B}\nonumber + 6 \bar{C}
      \nonumber \\ & &
      + 4 \bar{N}_e^{\mbox{\scriptsize
	num}}(0)\{3-2 \bar{A}^2-6\bar{B}\}
      \nonumber \\
0 &=& \omega^4 [4+19m(m\! -\!1)] + 10 \omega^2 (2m-1)[\bar{A}-6
	\bar{N}_e^{\mbox{\scriptsize num}}(0)]
      \nonumber \\ & &
      +180(\bar{N}_e^{\mbox{\scriptsize
	num}}(0))^2 -60\bar{A} \bar{N}_e^{\mbox{\scriptsize
	num}}(0) -6 +4\bar{A}^2 +12\bar{B}\; .
      \nonumber
\end{eqnarray}
The function cn$(z|m)$ is one of the Jacobian elliptic functions, and
$K(m)$ is the complete elliptic integral \cite{abramo64}.
If $m$ is close to one cn$(z|m)$ becomes almost identical to
the expression $1/\cosh (z)$ except that cn$^2(z|m)$ is periodic
in $z$ with period $2K(m) \approx \ln [16/(1-m)]$.
For a number state $|\psi_n \rangle$ we have $\bar{A}=4n+1$, $\bar{B}= n^2$,
and $\bar{C} = 2n^2 -n$. The corresponding solution (\ref{lsg}) is, to the
first two leading orders in $n \gg 1$, characterized by
$\omega \approx \sqrt{n+2}$ and $m \approx 1-2/n$. A second
exact solution of Eq.~(\ref{nosys}) is structurally
similar to that of Eq.~(\ref{lsg}) if cn$(z|m)$ is replaced by the elliptic
function dn$(z|m)$. In this case the parameters are determined by different
algebraic equations which we will omit in this paper.

\begin{figure}[t]
\epsfysize=5.8cm
\hspace{-2mm}
\epsffile{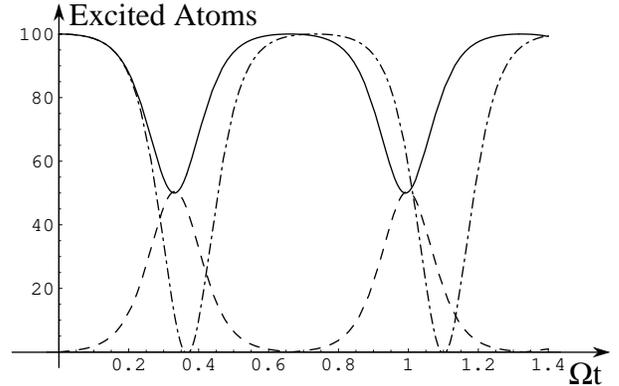}%\vspace{-1cm}
\caption{Mean value $\bar{N}_e^{\mbox{\scriptsize num}}$
(solid curve) and variance $\Delta_e$ (dashed curve) of the number of excited
atoms as a function of time for 100 initially excited atoms in an atomic
number state. The dot-dashed curve is obtained in the vanishing
variance approximation. $\Omega$ is the generalized Rabi frequency.}
\end{figure}
Our improved mean value solution $\bar{N}_e^{\mbox{\scriptsize num}}(\tau)$
of Eq.~(\ref{lsg}) is plotted in Fig.~1 together with the ordinary mean value
solution obtained in the vanishing variance approximation ($\Delta_e =0$).
As one can see, the fluctuations become very large when the dip in the
number of excited atoms occurs.
As compared with the vanishing-variance approximation
the dip is reduced by a factor of 1/2.
This is in qualitative agreement with the results of Ref.~\cite{haake70}
where a similar physical system was examined under the assumption that the
photons escape quickly so that the number of excitations is not conserved.
Note that the frequency of the occurrence of the dip is modified, too.
The period $T_p$ is
determined by $\sqrt{n+2} \Omega T_p = 2 K(1-2/n)$ leading to $T_p = \ln
(8n)/[\sqrt{n+2} \Omega]$. It diminishes with increasing number of atoms.
%%%%%%%%%%%%%%%%%%%%%%%%%%%%%%%%%%%%%%%%%%%%%%%%%%%%%%%%%%%%%%%%%%%%
\section{Collapse and revival}
Based on the solution (\ref{lsg}) we are now able to calculate the mean
number of excited atoms if initially no photons and
only excited atoms are present which
form a coherent state with mean number $\bar{n}$ of atoms.
We expand this state in the number states
$|\psi_l \rangle$. Since each of the states $
\sqrt{N_e(0)}\exp [-i \tilde{H}t/\hbar]
|\psi_l \rangle$ is an eigenstate of $S_A$ with eigenvalue $l$ they are
orthogonal for different $l$. This reduces the mean value of $N_e$ to
the expression
\begin{equation}
\bar{N}_e^{\mbox{\scriptsize coh}} (\tau)  = e^{- \bar{n}}
\sum_{l=0}^\infty \frac{\bar{n}^{l}}{l!} \langle \psi_l | N_e (\tau)
| \psi_l \rangle
\label{coh} \end{equation}
Because the r.h.s.~is only a function of the mean values $\bar{N}_e
^{\mbox{\scriptsize num}}(\tau)$ of the number state case, it may
directly be evaluated with the help of Eq.~(\ref{lsg}).
The physical reason for this simplification is that the interaction conserves
the total number of atoms. Thus, no interferences between number
states of different total atom number can occur. Note that the
expression (\ref{coh}) is identical to the mean value for a statistical
mixture of number states with Poissonian statistics.

\begin{figure}[t]
\epsfysize=6.1cm
\hspace{-4mm}
\epsffile{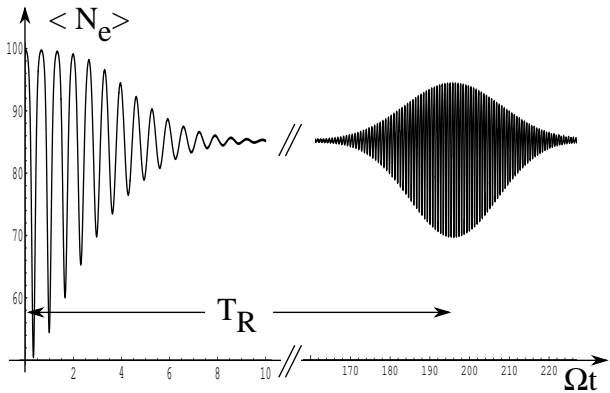}
%\vspace{-0.5cm}
\caption{For a coherent state of 100 initially excited atoms or a Poissonian
mixture of number states the time evolution of the mean number of excited
atoms exhibits collapses and revivals.}
\end{figure}
Whereas $\bar{N}_e^{\mbox{\scriptsize num}}(\tau)$ for a number state shows a
periodic behaviour with
period $T_p$ it is an important consequence of Eq.~(\ref{coh}) that a
coherent atomic state as initial state
leads to collapse and revival in the number $\bar{N}_e^{\mbox{\scriptsize
coh}}(\tau)$ of excited atoms (see Fig.~2). This is also the case for a
statistical mixture. Leaving open the question if a condensate can at all
be described by a coherent state (compare, e.g., Ref.~\cite{zoller96}) an
experimental verification of collapse and revival cannot be used to
discriminate between coherent state and mixture.

We turn to the details of the time development of
$\bar{N}_e^{\mbox{\scriptsize coh}} (\tau)$.
The time $T_R$ when the first (large) revival occurs can be determined by
the criterion given in Ref.~\cite{eberly80}:
the expression $\langle \psi_l | N_e (\tau) | \psi_l \rangle$
must simultaneously have a
maximum for the two neighboring values $l=\bar{n}$ and
$l=\bar{n}+1$. This guarantees that many terms in the sum
in Eq.~(\ref{coh}) become simultaneously large.
Since cn$(z|m)$ is periodic this condition becomes a condition
on the argument of the Jacobian elliptic function, namely
$\sqrt{\bar{n}} \Omega T_R -K(1-2/\bar{n}) = 2 r K(1-2/\bar{n})$ and
$\sqrt{\bar{n}+1} \Omega T_R -K(1-2/(\bar{n}+1)) = 2 (r+1)
K(1-2/(\bar{n}+1))$ with an integer $r$. This can be solved for $r$ and $T_R$
and results to leading order in $\bar{n}$ in
\begin{equation}
T_R \approx {2\sqrt{\bar{n}} \over \Omega}
   \frac{\ln^2(8 \bar{n})}{\ln (8 \bar{n})-2}\; .
\end{equation}
According to its derivation the revival time $T_R$ is independent of the
probability weights of the mixture or state when it is expanded in $|\psi_l
\rangle$. It is therefore the same for all mixtures with the same initial
mean number of excited atoms.

In addition to the large revival at time $T_R$ there are many smaller
revivals at earlier times which arise at fractions of $T_R$
(see Fig.~3). These fractional revivals
have been predicted numerically for the degenerate parametric oscillator
in Refs.~\cite{iyyanki97,drobny92}. The fractional revivals arise when only a
part of the sum terms in Eq.~(\ref{coh}) have simultaneously a maximum.
Their revival times can be calculated by assuming that the terms for
$l= \bar{n}$ and $l= \bar{n}+ r, \; r=2,3,\ldots$ are simultaneously maximal.
This leads to $T_R (r) = T_R(1)/r$. Since the revivals reappear periodically
one generally finds them at fractions of small integers of the time $T_R$.
\begin{figure}[t]
\epsfysize=6.1cm
\hspace{-7mm}
\epsffile{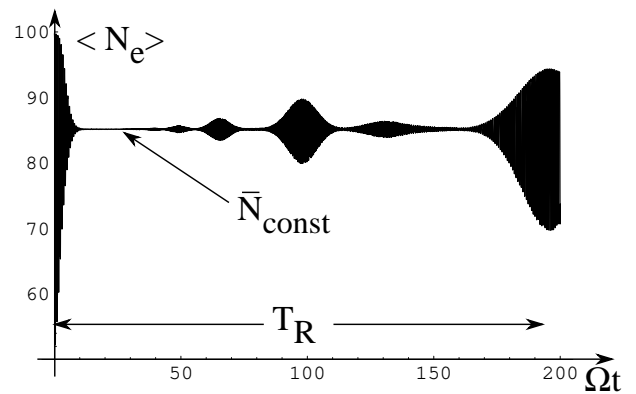}
%\vspace{-0.5cm}
\caption{Between the collapse and the main revival fractional revivals do
appear.}
\end{figure}

An estimation of the almost constant value
$\bar{N}_{\mbox{\scriptsize const}}$ taken by $\bar{N}_e
^{\mbox{\scriptsize coh}} (\tau)$ between the collapse and the first
revival can be derived under the assumption that the arguments of cn$(z|m)$
of Eq.~(\ref{lsg}) that appear
in the sum in Eq.~(\ref{coh}) are statistically distributed over the whole
period of cn$(z|m)$ (i.e., $z \in [0, 2K(m)]$).
$N_{\mbox{\scriptsize const}}$ then is simply the average of Eq.~(\ref{lsg})
over one period $T_p$. The corresponding integration
$(2K)^{-1} \int_0^{2K} \mbox{cn}^2(z|m) dz$
can be performed
if  cn$^2(z|m\approx 1)$ is approximated by $1/\cosh^2(z)$ and results to
leading order in $\bar{n}$ in
\begin{equation}
\bar{N}_{\mbox{\scriptsize const}} \approx \bar{n}\left \{ 1 - \frac{1}{\ln (
  8 \bar{n})} \right \} \; .
\end{equation}
The two analytical expressions for
$T_R$ and $N_{\mbox{\scriptsize const}}$ are in good agreement with
the numerical evaluation of Eq.~(\ref{coh}).

In conclusion we have presented analytical results for the
collapse and revival of the mean occupation number and the
variance of excited atoms in a microwave cavity. The
phenomenon of a revival in the context of Bose-Einstein condensation
was also examined for different physical systems. Zhang and Walls
\cite{zhang95} discovered it numerically for atoms passing
through a standing light wave. Wright {\em et al.} \cite{wright96}
showed that it occurs if the Bose condensate is described by a
superposition of states with different total number of atoms.
Our results differ from these approaches in that a third mode
(the microwave photons) is fully incorporated and that the
new theoretical method applied here allows to derive closed
expressions for the interesting physical quantities.
%%%%%%%%%%%%%%%%%%%%%%%%%%%%%%%%%%%%%%%%%%%%%%%%%%%%%%%%%%%%%%%%
{\bf Acknowledgement}: We thank Markus Holzmann for discussions
and the Optik-Zentrum Konstanz for financial support.
%%%%%%%%%%%%%%%%%%%%%%%%%%%%%%%%%%%%%%%%%%%%%%%%%%%%%%%%%%%%%%%%
%\newpage

\end{document}